\documentclass{emulateapj}
\usepackage{apjfonts}

\newcommand{\etal}{et~al.}
\newcommand{\MgIIdblt}{{\rm Mg}\kern 0.1em{\sc ii}~$\lambda\lambda 2796, 2803$}
\newcommand{\CaII}{\hbox{{\rm Ca}\kern 0.1em{\sc ii}}}
\newcommand{\CIV}{\hbox{{\rm C}\kern 0.1em{\sc iv}}}
\newcommand{\CV}{\hbox{{\rm C}\kern 0.1em{\sc v}}}
\newcommand{\MgI}{\hbox{{\rm Mg}\kern 0.1em{\sc i}}}
\newcommand{\MgII}{\hbox{{\rm Mg}\kern 0.1em{\sc ii}}}

\slugcomment{Accepted to AJ Oct. 29 2007}

\shorttitle{\sc {\MgII} Halo Sizes \& Gas Covering Fractions}
\shortauthors{\sc Kacprzak {\etal}}

\begin{document}


\title{Halo Gas Cross Sections And Covering Fractions of {\MgII}
Absorption Selected Galaxies}


\author{\sc
Glenn G. Kacprzak\altaffilmark{1},
Christopher W. Churchill\altaffilmark{1},
Charles C. Steidel\altaffilmark{2}, \\
and
Michael T. Murphy\altaffilmark{3,4}
}
                                                                                
\altaffiltext{1}{New Mexico State University, Las Cruces, NM 88003
{\tt glennk@nmsu.edu, cwc@nmsu.edu}}
 
\altaffiltext{2}{Caltech, Pasadena, CA 91125
{\tt ccs@astro.caltech.edu}}
 
\altaffiltext{3}{Institute of Astronomy, Cambridge CB3 0HA, UK}
 
\altaffiltext{4}{Swinburne University of Technology, Hawthorn,
Victoria 3122, Australia {\tt mmurphy@astro.swin.edu.au}}
\begin{abstract}

We examine halo gas cross sections and covering fractions, $f_c$, of
intermediate redshift {\MgII} absorption selected galaxies.  We
computed statistical absorber halo radii, $R_{\rm x}$, using current
values of $dN/dz$ and Schechter luminosity function parameters, and
have compared these values to the distribution of impact parameters
and luminosities from a sample of 37 galaxies.  For equivalent widths
$W_r(2796) \geq 0.3$~{\AA}, we find $43 \leq R_{\rm x} \leq 88$~kpc,
depending on the lower luminosity cutoff and the slope, $\beta$, of
the Holmberg--like luminosity scaling, $R \propto L^{\beta}$.  The
observed distribution of impact parameters, $D$, are such that several
absorbing galaxies lie at $D>R_{\rm x}$ and several non--absorbing
galaxies lie at $D < R_{\rm x}$.  We deduced $f_c$ must be less than
unity and obtain a mean of $\left< f_c \right> \sim 0.5$ for our
sample. Moreover, the data suggest halo radii of {\MgII} absorbing
galaxies do not follow a luminosity scaling with $\beta$ in the range
of $0.2-0.28$, if $f_c= 1$ as previously reported. However, provided
$f_c \sim 0.5$, we find that halo radii can remain consistent with a
Holmberg--like luminosity relation with $\beta \simeq 0.2$ and $R_{\ast}
= R_{\rm x}/\sqrt{f_c} \sim 110$~kpc.  No luminosity scaling
($\beta=0$) is also consistent with the observed distribution of
impact parameters if $f_c \leq 0.37$. The data support a scenario in
which gaseous halos are patchy and likely have non--symmetric
geometric distributions about the galaxies.  We suggest halo gas
distributions may not be govern primarily by galaxy mass/luminosity
but also by stochastic processes local to the galaxy.

\end{abstract}




\section{Introduction}

Understanding galaxy formation and evolution is one of the most
important topics of modern astronomy. The extended distribution of
baryonic gas surrounding galaxies holds great potential for
constraining theories of their formation. However, the sizes of
gaseous galaxy halos along with the distribution of gas within are not
well understood. Numerical models have been able to synthesize the
formation and evolution of large scale structures, however, there are
unresolved issues regarding the evolution of individual galaxies and
halos. The halo baryon--fraction problem \citep[e.g.,][]{mo02} and the
rapid cooling of gas \citep[e.g.,][]{white78} result in galaxy halos
which have little or no gas soon after they form. These effects are
not seen in the observable universe since there is an abundance of
galaxies where gas has been detected in halos via quasar absorption
lines.

From an observational standpoint, quasar absorption lines provides a
unique means of probing the extent and abundance of halo gas.
Although, quasar absorption line observations to date are sufficient
the recognize the aforementioned problems, they are lacking the detail
required to statistically constrain the distribution of the baryonic
gas in the halos of simulated galaxies. Cross--correlations between
absorbers and galaxies hold the promise to yield useful information on
cloud sizes and halo gas covering fractions. First steps towards
incorporating multi--phase gas in semi--analytical models and
numerical simulations suggest that warm gas in halos extends out to
galactocentric distances of $\sim 150$~kpc with cloud covering
fractions of $\sim 0.25-0.6$ \citep{maller04,kaufmann06}.

The association of {\MgIIdblt} doublet absorption in quasar spectra
with normal, bright, field galaxies has been firmly established
\citep[e.g.,][]{bb91,sdp94,cwc-china}.  In an effort to understand
halo sizes and gas distributions, \citet[][hereafter S95]{steidel95}
searched for foreground galaxies associated with {\MgII} absorption
within $\sim10''$ ($\sim65$~kpc for $z=0.5$) of
quasars\footnote{Throughout we adopt a $h=0.70$, $\Omega_{\rm M}=0.3$,
$\Omega_{\Lambda}=0.7$ cosmology. All quoted physical quantities from
previously published works have been converted to this cosmology.}.
The sample consisted of 53 absorbing and 14 non--absorbing galaxies
with a {\MgII} $\lambda 2796$ equivalent width sensitivity limit of
$W_{r}(2796) > 0.3$~{\AA}. S95 directly fitted the data by assuming a
Holmberg--like luminosity scaling,
\begin{equation}
R(L) = R_{\ast} \left( \frac{L}{L^{\ast}} \right) ^{\beta} \quad {\rm kpc},
\label{eq:rl}
\end{equation}
and minimizing the number of non--absorbing and absorbing galaxies
above and below the $R(L)$ relation. The best fit obtained clearly
showed that absorbing and non--absorbing galaxies could be separated
and that the halo radii $R(L_K)$ and $R(L_B)$ scale with luminosity
with $\beta = 0.15$ and $\beta = 0.2$, respectively, where an
$L_B^{\ast}$ galaxy has a gas halo cross section of $R_{\ast} =
55$~kpc. Furthermore, since almost none of the absorbing galaxies were
observed above the $R(L)$ boundary and that almost none of the
non--absorbing galaxies were observed below the $R(L)$ boundary, S95
inferred that {\it all\/} $L> 0.05L^{\ast}$ galaxies are hosts to
{\MgII} absorbing gas halos characterized by a covering fraction of
unity and a spherical geometry which truncates at $R(L)$. Examination
of this now ``standard model'' has been the subject of several
theoretical studies \citep[e.g.,][]{cc96,mo96,lin01}.

\citet{gb97} determined a steeper value of $\beta = 0.28$ for the
B--band luminosity obtained from a best fit to the upper envelope of
the distribution of impact parameters of 26 absorbing galaxies. They
found $R_{\ast} = 67$~kpc.

Using a reverse approach of establishing foreground galaxy redshifts
and then searching for {\MgII} absorption in the spectra of background
quasars yields results inconsistent with a covering fraction of
unity. For example, \citet{bowen95} identified 17 low--redshift
galaxies with background quasar probing an impact parameter range
between $3-162$~kpc. Galaxies that were probed at impact parameters
greater than 13~kpc had no absorption in the halo ($W_{r}(2796) \geq
0.40-0.9$~{\AA}), however, four of the six galaxies within 13~kpc of
the halo produced {\MgII} absorption. For intermediate redshift
galaxies, \citet{bechtold92} reported a covering fraction $f_c \simeq
0.25$ for $W_{r}(2796) \geq 0.26$~{\AA} for eight galaxies with $D
\leq 85$ kpc. Also, \citet{tripp-china} reported $f_c \sim 0.5$ for
$W_{r}(2796) \geq 0.15$~{\AA} for $\sim 20$ galaxies with $D \leq 50$
kpc.  These results are also consistent with the findings of
\citet{cwc-china} who reported very weak {\MgII} absorption,
$W_{r}(2796) < 0.3$~{\AA}, well inside the $R(L)$ boundary of bright
galaxies; these galaxies would be classified as ``non--absorbers'' in
previous surveys.  They also report $W_r(2796) > 1$~{\AA} absorption
out to $\simeq 2 R(L)$.  All these results suggest that there are
departures from the standard model, that the covering fraction of
{\MgII} absorbing gas is less than unity, and that the halo sizes and
the distribution of the gas appear to diverge from the $R(L)$ relation
with spherical geometry.

Another approach to understanding halo sizes and gas distributions is
to determine the statistical properties of {\MgII} absorbing gas and
then compute the statistical cross section from the redshift path
density, $dN/dz$ \citep[see][]{lanzetta95}.  The downfall of this
method is that a galaxy luminosity function must be adopted in order
to estimate $R_{\ast}$.  \citet{nestor05} acquired a sample of over
1300 {\MgII} absorption systems, with $W_{r}(2796) \geq 0.3$~{\AA}
from the Sloan Digital Sky Survey (SDSS). Using the $K$--band
Holmberg--like luminosity scaling and luminosity function of MUNICS
\citep{drory03}, Nestor {\etal} computed $R_{\ast} = 60-100$~kpc for
adopted minimum luminosity cutoffs of
$L_{min}=0.001-0.25L^{\ast}$. They found no redshift evolution of
$R_{\ast}$ over the explored range of $0.3\leq z \leq 1.2$.

\citet{zibetti06} studied the statistical photometric properties of
$\sim2800$ {\MgII} absorbers in quasar fields imaged with SDSS.  Using
the method of image stacking, they detected low--level surface
brightness (SB) azimuthally about the quasar. The SB profiles follow a
decreasing power law with projected distance away from the quasar out
to $100-200$~kpc. These results imply that absorption selected
galaxies may reside out to projected distances of 200~kpc.  However,
it is worth noting that the extended light profiles may be an artifact
of clustering of galaxies. Cluster companions of the {\MgII} absorbing
galaxies could extend the observed light profile over hundreds of
stacked images. Thus, one would infer that {\MgII} absorbing galaxies
are present at a larger impact parameters than would be found in
direct observation of individual galaxies.


Motivated by recent expectations from simulations that halo gas is
dynamically complex and sensitive to the physics of galaxy formation,
we investigate the standard halo model of {\MgII} absorbers. We also
aim to provide updated constraints on $f_c$ and $\beta$ for galaxy
formation simulations. In this paper, we demonstrate that $f_c < 1$
and question the validity of the Holmberg--like luminosity scaling
(Eq.~\ref{eq:rl}). Using high resolution quasar spectra, we explore
{\MgII} absorption strengths to an order of magnitude more sensitive
than previous surveys which allow us to re--identify non--absorbing
galaxies as ``weak'' absorbing galaxies.  In \S~\ref{sec:data} we
describe our sample and analysis. In \S~\ref{sec:results}, we present
new calculations of the statistical absorber radius computed using the
statistically measured absorption path density $dN/dz$ and the
Schechter luminosity function.  We then compare these values to the
empirical results of S95 and to a sample of known {\MgII} absorption
selected galaxies with measured luminosities and impact parameters. We
also examine how individual halos behave with respect to the
statistical halo. In \S~\ref{sec:dis}, we discuss the properties and
distribution of gas in halos.  Our concluding remarks are in
\S~\ref{sec:conclusion}.

\section{Data and Analysis}
\label{sec:data}

\begin{figure*}[htb]
\includegraphics[angle=0,scale=0.80]{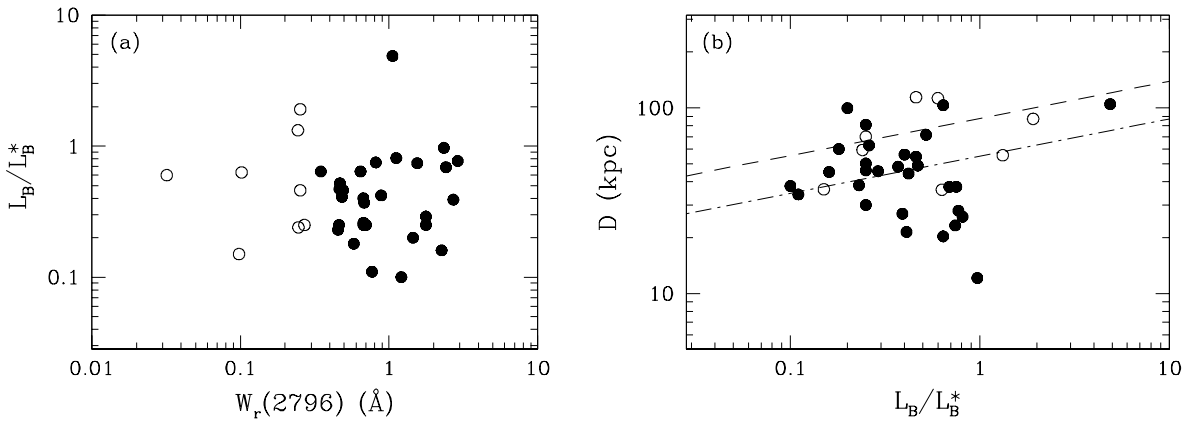}
\caption{ (a) $W_r(2796)$ versus $L_B/L_B^{\ast}$. Filled circles have
$W_r(2796)\geq 0.3$~{\AA} and the open circles have
$W_r(2796)<0.3$~{\AA}. --- (b) The impact parameter, $D$, versus
$L_B/L_B^{\ast}$.  The dash--dot line is the halo luminosity scaling
given by Eq.~\ref{eq:rl} for the results of S95
($R_{\ast}=55$~kpc, $f_c=1$, $\beta = 0.2$). The dash--dash line is the
halo luminosity scaling given by Eq.~\ref{eq:rl} for our result
($R_{\ast}=88$~kpc), assuming $f_c=1$, $\beta = 0.2$.}
\label{fig:ewd}
\end{figure*}

We have constructed a sample of 37 galaxies ($0.3<z<1.0$), with
spectroscopically confirmed redshifts, selected by the presence of
{\MgII} absorption in quasar spectra. The absorption properties were
measured from HIRES/Keck \citep{vogt94} and UVES/VLT \citep{dekker00}
spectra. The {\MgII} $\lambda 2796$ profiles have been presented in
\citet{cwc-china}, where the detection limit is $W_r(2796)\geq
0.02$~{\AA} (5~$\sigma$).  Galaxy properties were measured from F702W
or F814W WFPC--2/{\it HST} images of the quasar fields. Images of the
galaxies, along with further details of the sample selection, data,
and data analysis, can be found in \citet{kacprzak07}.

Galaxy absolute magnitudes, $M_B$, were determined from the
$k$--corrected observed $m_{F702W}$ or $m_{F814W}$ adopted from
\citet{kacprzak07}.  The $k$--corrections were computed using the
formalism of \citet{kim96} based upon the spectral energy distribution
(SED) templates of \citet{kinney96}.  The adopted SED for each galaxy
was based upon its rest--frame $B-K$ color (SDP94). For galaxies with
no color information, we adopted a Sb SED which is consistent with
average color of {\MgII} absorbing galaxies
\citep[SDP94;][]{zibetti06}.  Our $k$--corrections are consistent with
those from the literature \citep{kim96,fukugita95}.  $B$--band
luminosities were computed using the DEEP2 optimal $M^{\ast}_B$ of
\citet[][Table~2]{faber05} in the redshift bin appropriate for each
galaxy.  $M^{\ast}_B$ ranges from $-21.07$ ($\left< z \right> =0.3$)
to $-21.54$ ($\left< z\right> =1.1$). 

We compute the halo gas cross section determined from the redshift
path density,
\begin{equation}
\frac{dN}{dz} = \pi R_{\rm x}^2 \cdot \Phi^{\ast}\Gamma(x,y) \cdot
\frac{c}{H_0}\frac{(1+z)^2}{\sqrt{\Omega_m(1+z)^3+\Omega_{\Lambda}}},
\label{eq:dndz}
\end{equation}
where $R_{\rm x}$ is the statistical absorber radius for an $L^{\ast}$
galaxy, and $\Phi^{\ast}$ is the number density of $L^{\ast}$
galaxies.  $R^2_{\rm x}=f_c R^2_{\ast}$, where $R_{\ast}$ is the
covering fraction corrected absorbing halo radius.  Note here that we
make a distinction between $R_{\rm x}$, which is derived from the
redshift path density, and $R_{\ast}$, which is a physical cross
section of the absorbing gas accounting for the covering fraction.
$\Gamma(x,y)$ is the incomplete Gamma function in which $x=2\beta -
\alpha +1$, where $\alpha$ is the faint--end slope of the Schechter
galaxy luminosity function and $\beta$ parameterizes a Holmberg--like
luminosity scaling of Eq.~\ref{eq:rl}. The parameter
$y=L_{min}/L^{\ast}$, where $L_{min}$ is the minimum luminosity of
galaxies contributing to absorption.  The influence of $y$ on the
value of $R_{\rm x}$ becomes relatively more important as $\beta
\rightarrow 0$.

We present our sample in Figure~\ref{fig:ewd}$a$, plotting $W_r(2796)$
versus $L_B/L^{\ast}_B$. The solid points have $W_r(2796)\geq
0.3$~{\AA} and the open points are weak systems \citep{weakI}, having
$W_r(2796)<0.3$~{\AA}, and would have been classified as
non--absorbing galaxies in previous surveys \citep[e.g.,
SDP94;][]{gb97}. {\it Since $R_{\rm x}$ is computed using $dN/dz$
which is determined for systems with} $W_{r}(2796) \geq 0.3$~{\AA},
{\it we must consider these ``weak'' systems as ``non--absorbing''
galaxies in order to be consistent with our comparisons for the
remainder of this paper}. In Figure~\ref{fig:ewd}$a$ note that both
absorbing and non--absorbing galaxies span the same luminosity range.

\section{Results}
\label{sec:results}

Applying Eq.~\ref{eq:dndz}, we computed the statistical absorption
radius, $R_{\rm x}$, for $W_r(2796) \geq 0.3$~{\AA} employing the most
current Schechter luminosity function parameters and absorber redshift
path density. We adopted $dN/dz=0.8$ \citep{nestor05}, $\alpha=1.3$,
and $\Phi_{\ast}=3.14\times 10^{-3}$ Gal Mpc$^{-3}$ \citep{faber05}
for the $\left< z \right> =0.5$ redshift bin, where the mean redshift
of our sample is 0.58.  Since the luminosity scaling is not
necessarily constrained by our sample, we consider both $\beta=0.2$
and $\beta=0$ (i.e., no scaling) for $y=0.05$ and $y=0.01$. We
obtained,
\begin{equation}
R_{\rm x}=\sqrt{f_c}R_{\ast} =  \left\{ 
\begin{array}{r@{~, \quad}l}
64\mbox{ kpc}  & y=0.05,~\beta=0 \\
43\mbox{ kpc}  & y=0.01,~\beta=0 \\
88\mbox{ kpc}  & y=0.05,~\beta=0.2 \\
72\mbox{ kpc}  & y=0.01,~\beta=0.2~. 
\end{array}
\right.
\end{equation}

By {\it direct fitting\/} of his sample, S95 empirically deduced
$R_{\ast}= 55$~kpc and inferred $f_c=1$, $\beta = 0.2$ and
$y=0.05$. Assuming $f_c=1$, $\beta = 0.2$ and $y=0.05$, we computed a
statistical covering fraction corrected absorber halo radius of
$R_{\ast}= 88$~kpc. The difference between the two values arises from
the different methods used to determine $R_{\ast}$; S95 applied a fit
to a known sample of {\MgII} absorption selected galaxies, whereas,
our values are directly computed from measured absorption and galaxy
statistics. Assuming $f_c$ less than unity would increase our computed
value of $R_{\ast}$, yielding a value even less consistent with that
of S95.

In Figure~\ref{fig:ewd}$b$, the projected quasar--galaxy separation,
$D$, is plotted versus $L_B/L^{\ast}_B$. The mean impact parameter is
$\left<D\right>=53.2$~kpc which is close to the S95 halo size. The
dash--dot line is the halo radius, $R(L)$, from Eq.~\ref{eq:rl} using
$R_{\ast}=55$~kpc, $f_c=1$, and $\beta = 0.2$ found by S95. Three
non--absorbing galaxies reside below the $R(L)$ boundary and five
reside above. This is not necessarily inconsistent with S95, who found
two of 14 non--absorbing galaxies below the $R(L)$ boundary. However,
we find 16 $W_r(2796) \geq 0.3$~{\AA} absorbers that are outside the
$R(L)$ boundary by as much as 60~kpc. In the standard halo model,
galaxies above the $R(L)$ boundary are expected to not be associated
with $W_r(2796) \geq 0.3$~{\AA} absorption.  The dash--dash line is
the halo radius, $R(L)$, from Eq.~\ref{eq:rl} using the parameters
$R_{\rm x}=88$~kpc, $f_c=1$, and $\beta = 0.2$.  We find that five of
the eight non--absorbing galaxies lie below the $R(L)$ boundary. These
five galaxies are expected to be strong absorbing galaxies if they obey
the $R(L)$ relation. Also, there are three absorbing galaxies above
the $R(L)$ boundary.

\begin{figure*}
\includegraphics[angle=0,scale=0.80]{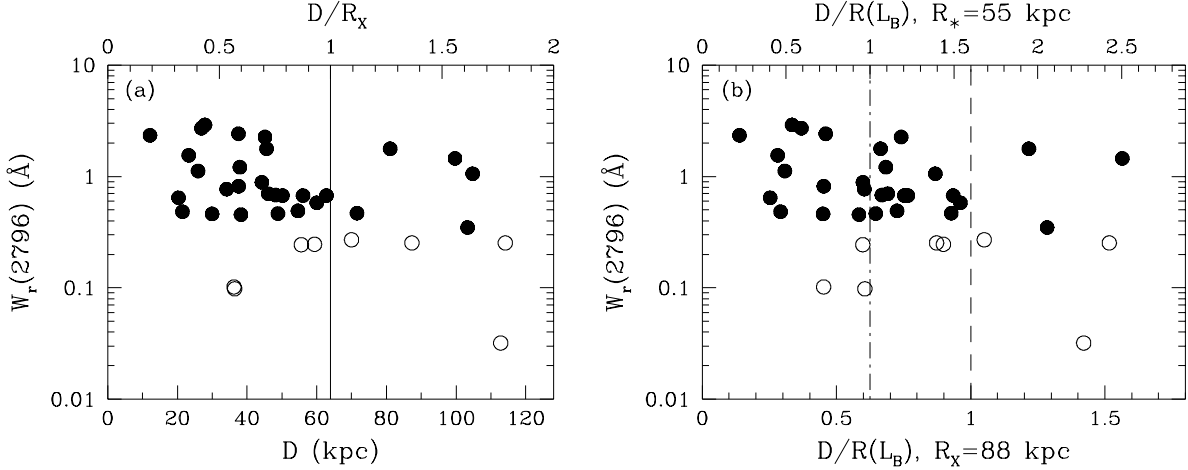}
\caption{ (a) $W_r(2796)$ as a function $D$ (bottom axis) and
$D/R_{\rm x}$ (top axis). The solid line represents $R_{\rm x}=64$~kpc
for $\beta = 0$ and for a luminosity cutoff of 0.05$L_B^{\ast}$. ---
(b) $W_r(2796)$ as a function of $D/R(L_B)$. The dashed--dotted line
represents $R_{\ast}$=55~kpc and the dash--dash represents $R_{\rm
x}=88$~kpc using $\beta = 0.2$ and $y=0.05$.}
\label{fig:l}
\end{figure*}

From Figure~\ref{fig:ewd}$b$ it would appear that the value of $\beta$
is not constrained for the B-band luminosities since non--absorbing
galaxies are both above and below $R(L)$ for both the
$R_{\ast}=55$~kpc deduced by S95 and our computed size of $R_{\rm
x}=88$~kpc.  Assuming that there is no luminosity scaling, we explore
halo cross sections with $\beta=0$.  In Figure~\ref{fig:l}$a$, we plot
$W_r(2796)$ versus $D$. The vertical line is the statistical absorber
radius, $R_{\rm x}=64$~kpc (where $D/R_{\rm x}=1$), for $\beta = 0$
and $y=0.05$. The top axis gives $D/R_{\rm x}$. Galaxies to the left
of the line are consistent with the computed statistical absorber
radius.  Galaxies to the right of the line are inconsistent; if the
standard halo model applies these particular galaxies must have halos
with $f_c<1$.  We find five of 29 galaxies at $D>R_{\rm x}$. If we
assume $y=0.01$ and $\beta=0$, we obtain $R_{\rm x}=43$~kpc and find
16 galaxies reside at $D>R_{\rm x}$ and that four non--absorbing
galaxies are expected to have $W_r(2796) \geq 0.3$~{\AA} absorption.
Note that $R_{\rm x}$ is very sensitive to the choice of the
luminosity cutoff when $\beta = 0$. Larger $\beta$ suppresses the
faint end slope in Eq.~\ref{eq:dndz}, reducing the cross sectional
contribution of the lowest luminosity galaxies that dominate by
number.

In Figure~\ref{fig:l}$b$, we plot $W_r(2796)$ versus $D/R(L)$. The
dash--dash line is $D/R(L)=1$ for $R_{\rm x}=88$~kpc, $\beta =0.2$ and
$y=0.05$. Again, three of 29 galaxies have $D/R(L)>1$ and five
non--absorbing galaxies have $D/R(L)<1$. If we assume $y=0.01$ and
$\beta=0.2$, we obtain $R_{\rm x}=72$~kpc, which increments the number
of galaxies at $D/R(L)>1$ to 7. If we apply $\beta=0.28$ from
\citet{gb97}, then $R_{\rm x}$ increases to 100~kpc (for $y=0.05$) and
only three absorbing galaxy lie above the $R(L)$ boundary. The dash--dot
line is the S95 result where $D/R(L)=1$ for $R_{\ast}=55$~kpc. It is
clear that there is a significant fraction of absorbing galaxies that
are not well represented by the standard halo model of S95, since it
is expected that spherically symmetric gas halos with unity covering
fraction would give rise to absorption exclusively at $D/R(L) \leq 1$.

\section{Discussion}
\label{sec:dis}

Our sample of galaxies is not statistically complete, due to the
chosen method of searching for galaxies selected by {\MgII} absorption
\citep[see][]{cwc-china}. None the less, the data clearly support a
covering fraction less than unity, based upon the deduced statistical
absorber radius, $R_{\rm x}$, which is computed from the redshift path
density of the full population of {\MgII} absorbers.

 As seen in Figures~\ref{fig:l}$a$ and \ref{fig:l}$b$, a substantial
fraction of galaxies are found at impact parameters well beyond the
statistical absorber radius. The largest impact parameter in our
sample is $D_{\rm max} \simeq 105$~kpc. If we assume that the largest
impact parameter is a proxy for the true size of the covering fraction
corrected absorbing halo radius such that $R_{\ast}=D_{\rm max}$ then
we can compute a luminosity function weighted covering fraction where
$f_c =w_L(\beta, y)(105/R_{\rm x})^{-2}$.  Assuming a lower galaxy
luminosity cutoff of $y=0.05$, we obtain $f_c = 0.37$ for $\beta=0$
and $f_c = 0.37$ for $\beta=0.2$. Assuming a lower luminosity cutoff
of $y=0.01$, we obtain $f_c = 0.17$ for $\beta=0$ and $f_c = 0.17$ for
$\beta=0.2$.  These results are summerized in Table~1. Now with
$f_c<1$, the presence of eight non--absorbing galaxies within the
statistical halo radius, $R(L)$, is consistent with $R_{\ast} \simeq
105$~kpc for $W_r(2796) \geq 0.3$~\AA.

\begin{deluxetable*}{ccccccc}[htb]
\tabletypesize{\footnotesize}
\tablecolumns{7} 
\tablewidth{0pt} 
\tablecaption{{\MgII} Halo Gas Convering Fractions\label{tab:table}}

\tablehead{ 
\colhead{ }  & &
\multicolumn{2}{c}{$y=0.05$} &&
\multicolumn{2}{c}{$y=0.01$} \\
\cline{3-4} \cline{6-7} 
&         &      &       &&        &     \\[-1.5ex]
\colhead{No.\tablenotemark{ a}} & 
\colhead{$f_c$} & 
\colhead{$\beta=0$, $R_{\rm x}=64$~kpc} & 
\colhead{$\beta=0.2$, $R_{\rm x}=88$~kpc}  & & 
\colhead{$\beta=0$, $R_{\rm x}=43$~kpc} &
\colhead{$\beta=0.2$, $R_{\rm x}=72$~kpc} 
}
\startdata 
1.&$w_L(\beta, y)(105/R_{\rm x})^{-2}$  & 0.37 &  0.37 && 0.17      & 0.17\\
                                        &      &       &&           &     \\[-2.0ex]
2.&$\left<D/R(L_B)\right>^{-2}$         & 0.52 &  0.79 && 0.58      & 0.63\\
                                        &      &       &&           &     \\[-2.0ex]
3.&$(R_{\rm x}/55)^{-2}$                & 0.74 &  0.40 &&$ \cdots $\tablenotemark{ b} & 0.58\\ 
                                        &      &       &&           &     \\[-2.0ex]\hline
                                        &      &       &&           &     \\[-1.5ex]
& $\left<f_c\right>$                    & 0.54 &  0.52 && 0.38      & 0.46\\[-6pt]
\enddata 

\tablenotetext{a}{The different methods for computed the covering
fractions: 1) -- The luminosity function weighted $f_c$, assuming
$R_{\ast}$ equal to the maximum impact parameter of $D_{\rm
max}=105$~kpc. 2) -- The average of the covering fractions for each
galaxy was computed for galaxies with impact parameters greater then
the statistical halo size. 3) -- The statistical halo size is assumed
to be 55~kpc (S95).}

\tablenotetext{b}{Our sample of galaxies provide no constraint on the
covering fraction for $y=0.01$ and $\beta=0$.}
\end{deluxetable*}


Using each galaxy from our sample, a conservative estimate of the
covering fraction is the mean of the upper limit on $f_c =
(D/R(L))^{-2}$. In a complete sample, each galaxy with $D>R(L)$ makes
a fractional contribution to reducing the gas covering fraction.
Galaxies with $D \leq R(L)$ provide no constraint.  If our sample is
representative of a complete sample, we obtain $\left<f_c\right>=0.52$
($y=0.05,\beta=0$), $\left<f_c\right>=0.79$ ($y=0.05,\beta=0.2$),
$\left<f_c\right>=0.58$ ($y=0.01, \beta=0$), and
$\left<f_c\right>=0.63$ ($y=0.01,\beta=0.2$).

If we assume $R_{\ast}=55$~kpc of S95 is the true halo size we can
also compute the covering fractions such that $f_c = (R_{\rm
x}/R_{\ast})^{-2}$. We obtain $f_c = 0.40$ ($y=0.05,\beta=0.2$), $f_c
= 0.58$, ($y=0.01,\beta=0.2$), and $f_c = 0.74$ ($y=0.05,
\beta=0$). $R_{\ast}>R_{\rm x}$ for $y=0.01,\beta=0$ yields no
constraint on $f_c$. The results of the above computations of the
covering fractions are summarized in Table~1.

From all the methods of estimating $f_c$, we obtain $\left<f_c\right>
\sim 0.5$ with a range of $0.17 \leq f_c \leq 0.80$. This is
consistent with $f_c=0.7-0.8$ deduced by \citet{cc96} from Monte Carlo
simulations of {\MgII} absorption selected galaxy surveys.  Our
average $f_c$ is also consistent with the result of
\citet{tripp-china}\footnote{\citet{tripp-china} have a $W_r(2796)\sim
0.1$~{\AA} (2$\sigma$) detection limit which translates to a
$W_r(2796)\sim 0.25$~{\AA} (5$\sigma$) detection limit as presented
here. We have removed all absorbers with $W_r(2796) < 0.25$~{\AA} in
order to compare our results at the same detection limit.} who find
$f_c \sim 0.55$ and higher than $f_c \sim 0.25$ determined by
\citet{bechtold92}. Also, \citet{cwc06} found a galaxy, probed well
within the $R(L)$ boundary, that exhibits no {\MgII} absorption to
$W_r(2796) \leq 7$~m{\AA}. All these results suggest $f_c<1$ for
{\MgII} absorbing gas with $W_r(2796) \geq 0.3$~{\AA}. Thus,
non--absorbing galaxies below the predicted halo size are expected.

Although the data do not clearly support a halo size--luminosity
scaling, if we apply $f_c \sim 0.5$ such that the covering fraction
corrected absorbing halo radius is $R_{\ast}=1.41R_{\rm x}$, a
Holmberg--like luminosity relationship with $\beta \simeq 0.2$ is not
ruled out for both $y=0.05$ and $y=0.01$.  We can further constrain
$f_c$, $R_{\ast}$, and $\beta$ with a maximum likelihood fit that
satisfies the distribution of impact parameters and luminosities of
our sample. In this analysis, $R_{\ast}=\sqrt{f_c}R_{\rm x}$ is a
function of $\beta$ as constrained by $dN/dz$. First, we assume that
all absorbing galaxies must reside below the $R(L)$ boundary. For
$y=0.05$, we find an upper limit of $f_c \leq 0.4$ for a range of $
0.02 \leq \beta \leq 0.24$ with $105 \leq R_{\ast} \leq 150$~kpc,
respectively. For lower covering fractions, the allowed ranges of
$\beta$ and $R_{\ast}$ increase. For $y=0.01$, we find $f_c \leq 0.2$
for $ 0.04 \leq \beta \leq 0.66$ with $110 \leq R_{\ast} \leq
290$~kpc, respectively . If we relax the condition such that one to
three absorption selected galaxies may reside above the $R(L)$
boundary, which could account for errors in the luminosities and/or
our finite sample, then the allowed ranges of $f_c$, $R_{\ast}$ and
$\beta$ increase. For these cases with $y=0.05$, we find an upper
limit of $f_c \leq 0.7$ for $ 0.18 \leq \beta \leq 0.58$ with $80 \leq
R_{\ast} \leq 150$~kpc, respectively. Thus, our sample is consistent
with a Holmberg--like luminosity relationship in the case $f_c \la
0.5$.

A central issue to this discussion is whether there is a fundamental
physical difference between the halos of non--absorbing (weak) and
absorbing galaxies or whether the difference in $W_r(2796)$ arise
only from a chance intersection of the quasar line of sight through a
single gas cloud or a gas cloud complex in these halos. Even if weak
$W_r(2796) < 0.3$~{\AA} systems are similar to strong systems, and
differ only by the number of clouds intersected along the quasar line
of sight, our arguments for constraining the halo gas covering
fraction for $W_r(2796) \geq 0.3$~{\AA} still hold.

Strong absorbers are typically characterized by a dominant and blended
subsystem and accompanied by significantly weaker subsystems at
relative velocities ranging from $\sim 40$ to $100$~km~s$^{-1}$
\citep{cv01}.  In fact, there may be different physical processes
governing the $W_r(2796)$ distribution of weak absorption associated
with strong absorbers and the general population of weak
absorbers. \citet{weakI} determined that the number density of the
general population of weak systems increase as $W_r(2796)$ decreases
down to 0.02~\AA.  Whereas, the {\MgII} equivalent width distribution
of intermediate-- and high--velocity subsystems in strong absorbers
turns over below $W_r(2796)\sim0.08$~{\AA} \citep[][]{cv01,mshar07}.
These facts suggest that lines of sight through galaxy halos often
probe a dominant, more massive structure surrounded by smaller
fragments of gas; a scenario consistent with patchy halos, in which
some lines of sight near galaxies would be expected to probe only
$W_r(2796)\ga 0.08$~{\AA} weak absorption.

There is also the possibility that some of the galaxies in our sample
having redshifts consistent with those of the {\MgII} absorbers may
not be the primary structure responsible for the absorption.  In some
cases there could be a faint unidentified galaxy located directly in
front of the quasar that cannot be identified even with careful
point--spread subtraction of the quasar \citep[see][]{s97}.  Thus, our
estimated values of $f_c$ and $R_{\ast}$ could be slightly skewed
toward smaller and larger values, respectively. It is difficult to
quantify this affect since such putative faint galaxies could actually
be companions to the galaxies in our sample.

\section{Conclusions}
\label{sec:conclusion}

In conclusion, the gas covering fraction must be less than unity since
the observed impact parameter distribution of absorbing galaxies does
not fall exclusively within the statistical absorber halo radius in
the range of $43 \leq R_{\rm x}\leq 88$~kpc.  The fact that some
absorbing galaxies are found at $D>R_{\rm x}$ and some non--absorbing
galaxies are found at $D<R_{\rm x}$ implies $f_c < 1$ and that the
standard halo model cannot describe halos on a case by case
basis. This highlights the power of using the statistics of absorption
line surveys to constrain the properties of halos in relation to the
measured distributions in absorption selected galaxy surveys.

By quantifying how individual galaxy halos deviate from a ``standard''
halo, we have obtained an average gas covering fraction of $\left< f_c
\right> \sim 0.5$. It is possible that $f_c$ exhibits both a radial
and an equivalent width dependence, though we cannot address this with
our sample.  Values of $f_c$ are likely to depend on galaxy star
formation rates, and galaxy--galaxy mergers and harassment histories;
processes that give rise to patchy and geometrically asymmetric gas
distributions. Alternatively, the absorption properties of
intermediate redshift halos may be governed by the dark matter over
density, $\Delta \rho /\rho$, and redshifts at which the galaxies
formed \citep{cwc06}.

Our results also show that, if $f_c<1$, the sizes of {\MgII} absorbing
halos can still follow a Holmberg--like luminosity relation with
$\beta$ in the range of $0.2-0.28$ \citep[S95;][]{gb97}, which
corresponds to $R_{\ast}\sim 110$~kpc. If $\beta=0$ is assumed, then
$f_c \leq 0.37$ for our sample to be consistent with no luminosity
scaling. In semi--analytical models in which {\MgII} absorbing gas is
infalling and is pressure confined within the cooling radius of hot
halos \citep[e.g.,][]{mo96,burkert00,lin00,maller04}, a Holmberg--like
luminosity relation in quasar absorption line systems naturally arises
\citep{mo96}. However, these models have great difficulty explaining
{\MgII} absorption at impact parameters greater than $\sim 70$~kpc.
If on the other hand halo gas spatial distributions are governed by
stochastic mechanical processes, as suggested by \citet{kacprzak07},
then there is no {\it a priori} reason to expect a clean halo--size
luminosity scaling. It is likely that some combination of these
scenarios contribute to the statistical values of $f_c$ and
$\beta$. Thus, it is reasonable to suggest that {\MgII} halos sizes
may not be strictly coupled to the host galaxy luminosity.

Further work on the cross--correlations between absorbers and galaxies
would provide better estimates of $f_c$ and $\beta$, two quantities
that provide direct constraints of galaxy formation simulations. Also
needed are additional constrains on the relative kinematics of the
absorbing halo gas and galaxies
\citep[e.g.,][]{s02,ellison03,kacprzak07b}. What is required is the
development of techniques to quantitatively compare observational data
with mock quasar absorption line analysis of simulated galaxy halos
\citep{cwcaas06}.

\acknowledgments Partial support from program \#10644 which was
provided by NASA through a grant from the Space Telescope Science
Institute. Partial support for G.G.K was also provided by Sigma--Xi
Grants in Aid of Research. G.G.K thanks NMSU for funding from the
Graduate Student Enhancement Grant.  M.T.M thanks PPARC for and
Advanced Fellowship.  We thank Hsiao-Wen Chen for discussions
regarding k-corrections. We also thank the anonymous referee for
insightful comments. Some of the data presented herein are based on
observations made with the NASA/ESA Hubble Space Telescope, obtained
from the data archive at the Space Telescope Institute.  STScI is
operated by the association of Universities for Research in Astronomy,
Inc. under the NASA contract NAS 5-26555.  Some spectroscopic data
were obtained at the W.M. Keck Observatory, which is operated as a
scientific partnership among the California Institute of Technology,
the University of California and NASA.  The Observatory was made
possible by the generous financial support of the W.M. Keck
Foundation.  Additional spectroscopic data are based on observations
made with European Southern Observatory Very Large Telescope at the
Paranal Observatories under various programs.





{\it Facilities:} \facility{HST (WFPC--2)}, \facility{Keck I (HIRES)},
\facility{VLT (UVES)}.

\end{document}